\documentclass[aps,pra,twocolumn,moshowpacs,floatfix,superscriptaddress]{revtex4-1}

\usepackage{dsfont} 
\usepackage{bm} 

\usepackage{bbold} 
\usepackage{xcolor}
\usepackage{times}
\usepackage{physics}
\usepackage{graphicx,braket}
\usepackage{hyperref}
\hypersetup{colorlinks=true,urlcolor=blue,linkcolor=blue,citecolor=blue}
\usepackage{amsmath,amssymb,amsfonts}
\usepackage{wrapfig}
\usepackage{siunitx}

\usepackage{cleveref}
\crefname{equation}{Eqs.}{Eqs.}
\Crefname{equation}{Equation}{Equations}
\crefrangelabelformat{equation}{(#3#1#4--#5#2#6)}
\crefmultiformat{equation}{Eqs. (#2#1#3}{, #2#1#3)}{#2#1#3}{#2#1#3}
\Crefmultiformat{equation}{Equations (#2#1#3}{, #2#1#3)}{#2#1#3}{#2#1#3}

\begin{document}
\title{Symmetries and Conservation Laws in Quantum Trajectories: Dissipative Freezing}

\author{Carlos S\'anchez Mu\~noz }
\email{carlos.sanchezmunoz@physics.ox.ac.uk}
\affiliation{Clarendon Laboratory, University of Oxford, Parks Road, Oxford OX1 3PU, United Kingdom}
\author{Berislav Bu\v ca}
\affiliation{Clarendon Laboratory, University of Oxford, Parks Road, Oxford OX1 3PU, United Kingdom}
\author{Joseph Tindall}
\affiliation{Clarendon Laboratory, University of Oxford, Parks Road, Oxford OX1 3PU, United Kingdom}
\author{Alejandro Gonz\'alez-Tudela}
\affiliation{Instituto de F\'isica Fundamental IFF-CSIC, Calle Serrano 113b, Madrid 28006, Spain}
\author{Dieter Jaksch}
\affiliation{Clarendon Laboratory, University of Oxford, Parks Road, Oxford OX1 3PU, United Kingdom}
\author{Diego Porras}
\affiliation{Instituto de F\'isica Fundamental IFF-CSIC, Calle Serrano 113b, Madrid 28006, Spain}

\newcommand{\down}{\op{g}{e}}
\newcommand{\up}{\op{e}{g}}
\newcommand{\downd}{\op{+}{-}} 
\newcommand{\upd}{\op{+}{-}}
\newcommand{\app}{a^\dagger}
\newcommand{\ssp}{\sigma^\dagger}
\newcommand*{\Resize}[2]{\resizebox{#1}{!}{$#2$}}%

\begin{abstract}
In driven-dissipative systems, the presence of a strong symmetry guarantees the existence of several steady states belonging to different symmetry sectors. Here we show that, when a system with a strong symmetry is initialized in a quantum superposition involving several of these sectors, each
individual stochastic trajectory will randomly select a single one of them and remain there for the rest of the evolution.
Since a strong symmetry implies a conservation law for the corresponding symmetry operator on the ensemble level, this selection of a single sector from an initial superposition entails a breakdown of this conservation law at the level of individual realizations.
Given that such a superposition is impossible in a classical, stochastic trajectory, this is a a purely quantum effect with no classical analogue.
Our results show that a system with a closed Liouvillian gap may exhibit, when monitored over a single run of an experiment, a behaviour completely opposite to the usual notion 
of dynamical phase coexistence and intermittency, which are typically considered hallmarks of a dissipative phase transition. We discuss our results with a simple, realistic model of squeezed superradiance. 
\end{abstract}
\date{\today} \maketitle

Driven dissipative systems are ubiquitous in many body physics and cavity QED ~\cite{amo09a,rodriguez17a,fink18a,carr13a,melo16a,fitzpatrick17a,baumann10a,
klinder15a,hamsen18a,teufel11a,kolkowitz12a,pigeau15a}. These systems are typically gapped and feature a unique, non-equilibrium steady state. In the regime of a dissipative phase transition (DPT), however, this gap vanishes and the null-space of the Liouvillian is spanned by several compatible steady-states.~\cite{kessler12a,minganti18a,carmichael15a,weimer15a,benito16a,
sieberer13a,sanchezmunoz18b,biondi17a,hwang18a,mendoza16a}. 
Due to their fundamental interest and practical applications, such as enhanced metrological properties~\cite{macieszczak16b,fernandezlorenzo17a}, DPTs have attracted a significant amount of attention, with much work being devoted to study the associated phenomena of bistability~\cite{fink18a,carr13a,melo16a,mendoza16a,letscher17a,muppalla18a,schuetz13a,schuetz14a}, hysteresis~\cite{rodriguez17a,hruby18a}, intermittency~\cite{lee12a,fitzpatrick17a,hruby18a,malossi14a,muppalla18a,ates12a}, multimodality~\cite{letscher17a,malossi14a}, metastability~\cite{macieszczak16a} and symmetry breaking~\cite{manzano14a,wilming17a,hannukainen18a}. All these effects are understood as different manifestations of the coexistence of several non-equilibrium phases. 
In particular, many experiments will look for intermittency as the hallmark of such phase coexistence~\cite{lee12a,fitzpatrick17a,hruby18a,malossi14a,muppalla18a,ates12a}. 
Intermittency is a phenomenon defined by a random switching between periods of high and low dynamical activity (for instance in the rate of photon emission). This behaviour, which is observed during a single run of the experiment,  is conveniently described using the formalism of quantum jumps in which the system is characterized in terms of a pure wavefunction that undergoes stochastic evolution~\cite{zoller87a,molmer92a,plenio98a}.

The timescale $\tau$ of this intermittency is given by the inverse of the Liouvillian gap or asymptotic decay rate (ADR), i.e. the eigenvalue $\lambda_2$ of the Liouvillian operator $\mathcal L$ with the second largest real part~\cite{macieszczak16b,flindt13a,hickey14a}. Since a DPT is defined by a vanishing Liouvillian gap~\cite{kessler12a,minganti18a}, it will necessarily imply that $\tau$ diverges. In most typical situations, this closing is reached in the thermodynamic limit of a many-body system.  Consequently, for any finite system, the long-time limit where intermittency is observable will, at least formally, exist.
There are, however, situations in which the Liouvillian gap vanishes exactly and such a long-time limit cannot be taken. This is the case of systems featuring a \emph{strong symmetry}~\cite{buca12a}. Liouvillians $\mathcal L$ with a strong symmetry have a degenerate steady state---implying that $\lambda_2=0$---and an associated conservation law for the symmetry operator, $\dot A=\mathcal L^\dagger A=0$~\cite{buca12a,albert14a}. Since the Liouvillian gap is closed exactly for any system size,  the long-time limit of intermittency described before does not exist, and the dynamics is split into different, unconnected ergodic symmetry sectors.

In this work, we study the quantum trajectories of open quantum systems with a strong symmetry. We show that, when initialized in a superposition involving different symmetry sectors, the system will evolve towards a single one of them in each individual trajectory,  remaining there for the rest of the realization. This non-ergodic phenomenon, that we term \emph{dissipative freezing}, is in stark contrast with the typical looked-for phenomenology of intermittency in a DPT and predicts a completely different dynamical behaviour at the level of individual realizations of the experiment. Related effects have already been discussed in different contexts: in Ref.~\cite{benoist17a} exponential stability of subspaces for quantum trajectories was demonstrated;  in Ref.~\cite{benoist14a} it was shown that a quantum stochastic master equation describing non-demolition measurements converges to a pure state; in~Ref.~\cite{vanHorssen14a} a similar effect was discussed for quantum Markov chains.
An important result of our work is to relate this phenomena to the symmetries of the master equation. Notably, it implies that the conservation law for the strong-symmetry operator is broken at the level of trajectories and can only be recovered under ensemble-averaging. This is a purely quantum phenomenon, since it requires an initial superposition of different symmetry sectors that cannot be implemented classically, i.e. a single classical trajectory is always fully realistic and located in only one of this sectors. Understanding this phenomena is important for the dynamical characterization of dissipative systems with a closed Liouvillian gap. This limit has been proven relevant in quantum metrology, since it yields a Heisenberg scaling (quadratic in time) of the quantum Fisher information~\cite{macieszczak16b}.

To discuss the effect of dissipative freezing, we analyse a model that can be solved numerically, yet displays a rich variety of non-ergodic dynamics. This model consists of a coherently-driven spin ensemble with squeezed, collective spin decay, which can be implemented by adiabatic elimination of a cavity mode coupled to a multi-component atomic condensate via cavity-assisted Raman transitions~\cite{jaksch01a,micheli03a,dimer07a,dallatorre13a,gonzaleztudela13b}.

\begin{figure}[t!]
\begin{center}
\includegraphics[width=1\columnwidth]{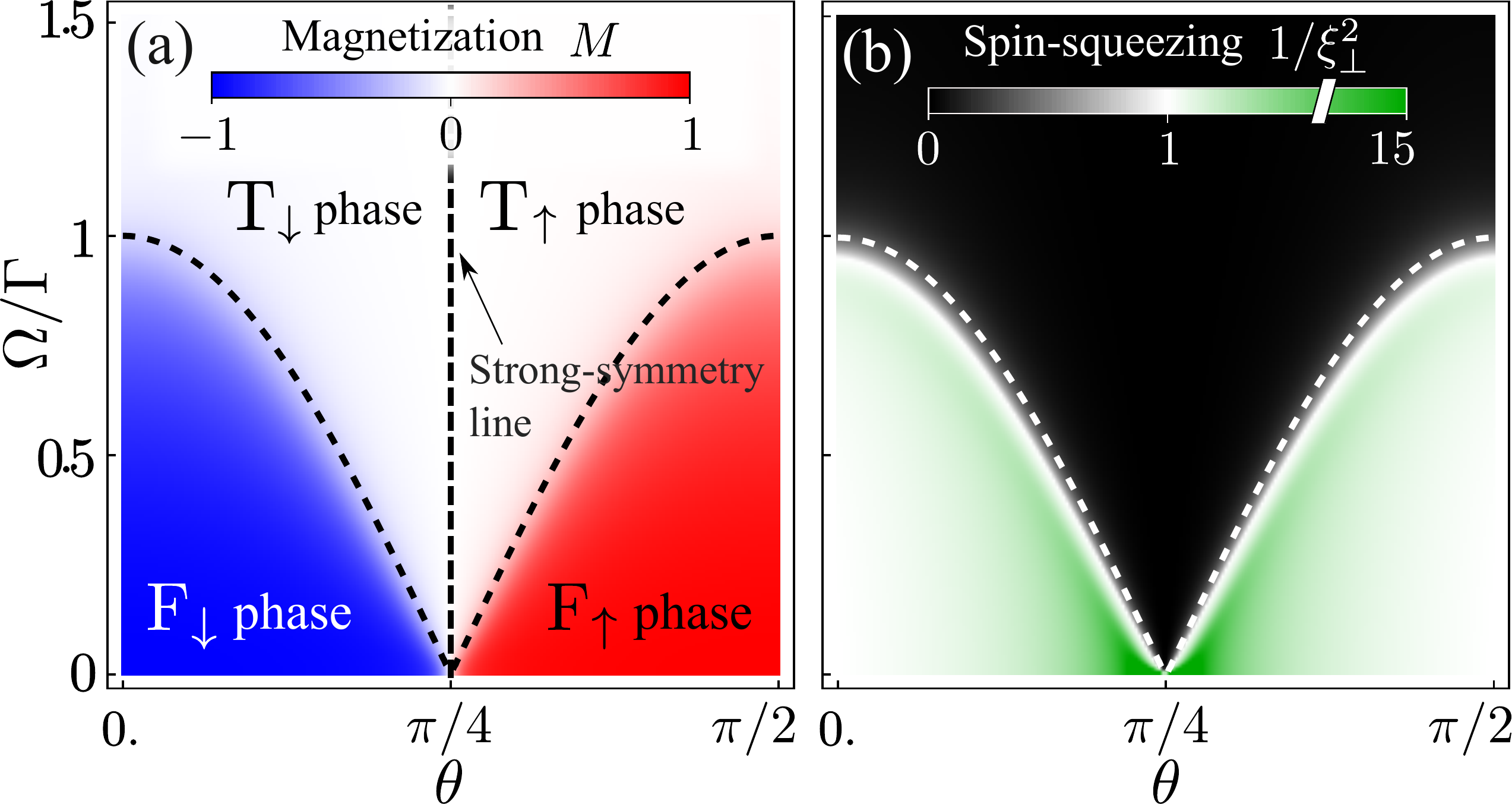}
\end{center}
\caption{Steady state observables for a finite system size $N=50$. Dashed lines indicate the critical line $\Omega_c(\theta)$. (a) Spin magnetization $M\equiv \langle S_z\rangle/J$, featuring ferromagnetic ($\text F$) and thermal ($\text T$) phases. (b) Spin squeezing $\xi_\bot^2\equiv N \langle\Delta S_x\rangle^2/\langle \mathbf S\rangle^2$.}
\label{fig:2}
\end{figure}

\emph{Model and phase diagram.---} The master equation describing the dynamics of the $N$-spin ensemble is ($\hbar=1$):
\begin{equation}
\dot{\rho}=-i\Omega[S_x,\rho]+\frac{\Gamma}{2J}\mathcal L_{D_\theta}[\rho],
\label{eq:master-equation}
\end{equation}
where $\mathcal{L}_O[\rho]\equiv 2 O\rho O^\dagger - \{O^\dagger O, \rho\} $ is the usual Lindblad superoperator~\cite{carmichael_book02a}, and the operator $D_\theta$ describes the quantum jumps undergone by the system, $D_\theta\equiv \cos(\theta) S_- + \sin(\theta) S_+$. In these equations, $S_\pm$ and $S_z$ are collective spin operators obeying angular momentum commutation relations, $\Omega$ is the driving amplitude, $\Gamma$ is the quantum-jump rate, and $J=N/2$ is the total angular momentum, which is conserved in the dynamics. The squeezed decay operator $D_\theta$ includes both $S_-$ and $S_+$, with a weight parametrized by the angle $\theta$ and an associated dark state which is a spin-squeezed state for $\theta\neq (0,\pi/2)$~\cite{dallatorre13a}.
Fig.~\ref{fig:2} depicts the phase diagram in the  $(\Omega,\theta)$ plane  in terms of the magnetization and spin-squeezing, featuring two types of non-equilibrium phases (discussed in more detail in Ref~\cite{arXiv_sanchezmunoz19a})
\emph{i)} The \emph{ferromagnetic (F) phase} is characterized by a well-defined magnetization,  diverging spin-squeezing at the phase transition, small fluctuations in the counting distributions of quantum jumps, high purity  and ergodic dynamics. Any initial state eventually relaxes into a stationary, almost pure gaussian steady-state. In the thermodynamic limit, this phase is well described within a Holstein-Primakoff approximation. 
\emph{ii)} In the \emph{thermal (T) phase} the steady-state is highly mixed, and close to the infinite-temperature state $\rho \propto \mathbb{1}$. This phase is characterized by zero mean magnetization,  small purity, large spin fluctuations, high rate of quantum jumps and large fluctuations in the output field.
The transition from the ferromagnetic to the thermal phase occurs at the critical driving $\Omega_c(\theta)=\Gamma(\cos^2\theta-\sin^2\theta)$. All the results presented here apply to the case $\theta\leq \pi/2$, trivially extended to $\theta \geq \pi/2$ by a spin flip. Hence, we have a spin-up and spin-down version of each phase, denoted as $\text F_{\uparrow/\downarrow}$ and $\text T_{\uparrow/\downarrow}$ in Fig.~\ref{fig:2}.

\emph{Liouvillian eigenvalues and symmetries.---} In the large driving limit of the thermal phase the Liouvillian features a particularly interesting spectrum of eigenvalues that can be derived analytically~\cite{arXiv_sanchezmunoz19a,ribeiro19a}:
\begin{equation}
\lambda_{q,k}^\pm=\pm iq\Omega-\frac{\Gamma_\theta}{2J}q^2-\frac{\chi_\theta}{4J}\left[q+k(1+k+2q)\right],
\label{eq:exact-eigenvalues}
\end{equation}
with $\Gamma_{\theta}\equiv \Gamma(\cos\theta+\sin\theta)^2$, $\chi_{\theta}\equiv \Gamma(\cos\theta-\sin\theta)^2$, $q=0,1,\ldots 2J$, $k = 0,1,\ldots 2J-q$. 
Equation~\eqref{eq:exact-eigenvalues} shows that, besides the steady-state eigenvalue $\lambda_{0,0}=0$,  other eigenvalues with zero real part can be obtained in two ways: either reaching the thermodynamic limit $J\rightarrow \infty$, or setting $\theta=\pi/4$ ($\chi_\theta=0$). For any fixed $q$,  $\lim_{J\rightarrow\infty}\Re[\lambda_{q,k}^\pm]=0$, which implies eigenstates with finite, purely imaginary eigenvalues and, therefore, the absence of stationarity and emergence of oscillatory dynamics in the long-time limit~\cite{buca19a}, which has recently attracted attention in similar models \cite{iemini18a,tucker18a}. The focus of this paper is, however, the situation $\theta=\pi/4$, where the Liouvillian gap closes exactly for any system size due to the presence of a \emph{strong symmetry}, i.e., an operator $A$ that satisfies $[H,A]=0$ and $[L_\mu,A]=0$, with $H$ the Hamiltonian and $L_\mu$ the set of quantum-jump operators of the master equation~\cite{buca12a}. In the model considered here, $A=S_x$. All the $\rho^{(m)}_0=|m\rangle\langle m|$ built from eigenstates $|m\rangle$ of $S_x$ are steady states of the dissipative dynamics~\footnote{Note that our system is fundamentally different from a single spin in a magnetic field pointing along the $x$ axis due to its driven-dissipative character: an initial superposition of different eigenstates of $S_x$ would never reach a stationary state in the purely Hamiltonian case.}.

\emph{Dissipative freezing of the dynamics.---}
The exact closing of the Liouvillian gap for any system size in the presence of a strong symmetry differs from the usual situation in which this closing, characteristic of a DPT~\cite{kessler12a,minganti18a,fink18a,fitzpatrick17a}, occurs in the thermodynamic limit~\cite{fink18a,carr13a,melo16a,mendoza16a,letscher17a,
lee12a,fitzpatrick17a,hruby18a,muppalla18a,ates12a,malossi14a}.
In the presence of a strong symmetry, multiple degenerate steady states can exist~\cite{buca12a} and the actual steady state of the system is then composed of a particular superposition of these states, fixed by the initial conditions~\cite{macieszczak16a,minganti18a}. Since in this case the evolution is not necessarily ergodic, it is not guaranteed that a single trajectory will switch among these states, which is the main assumption behind the notion of intermittency~\cite{lee12a,fitzpatrick17a,hruby18a,malossi14a,muppalla18a,ates12a}. More importantly, we want to clarify whether the conservation law $\dot A=0$ will apply at the level of an individual trajectory, since, to the best of our knowledge, this is only guaranteed when $A$ is unitary.

\begin{figure}[t!]
\begin{center}
\includegraphics[width=1\columnwidth]{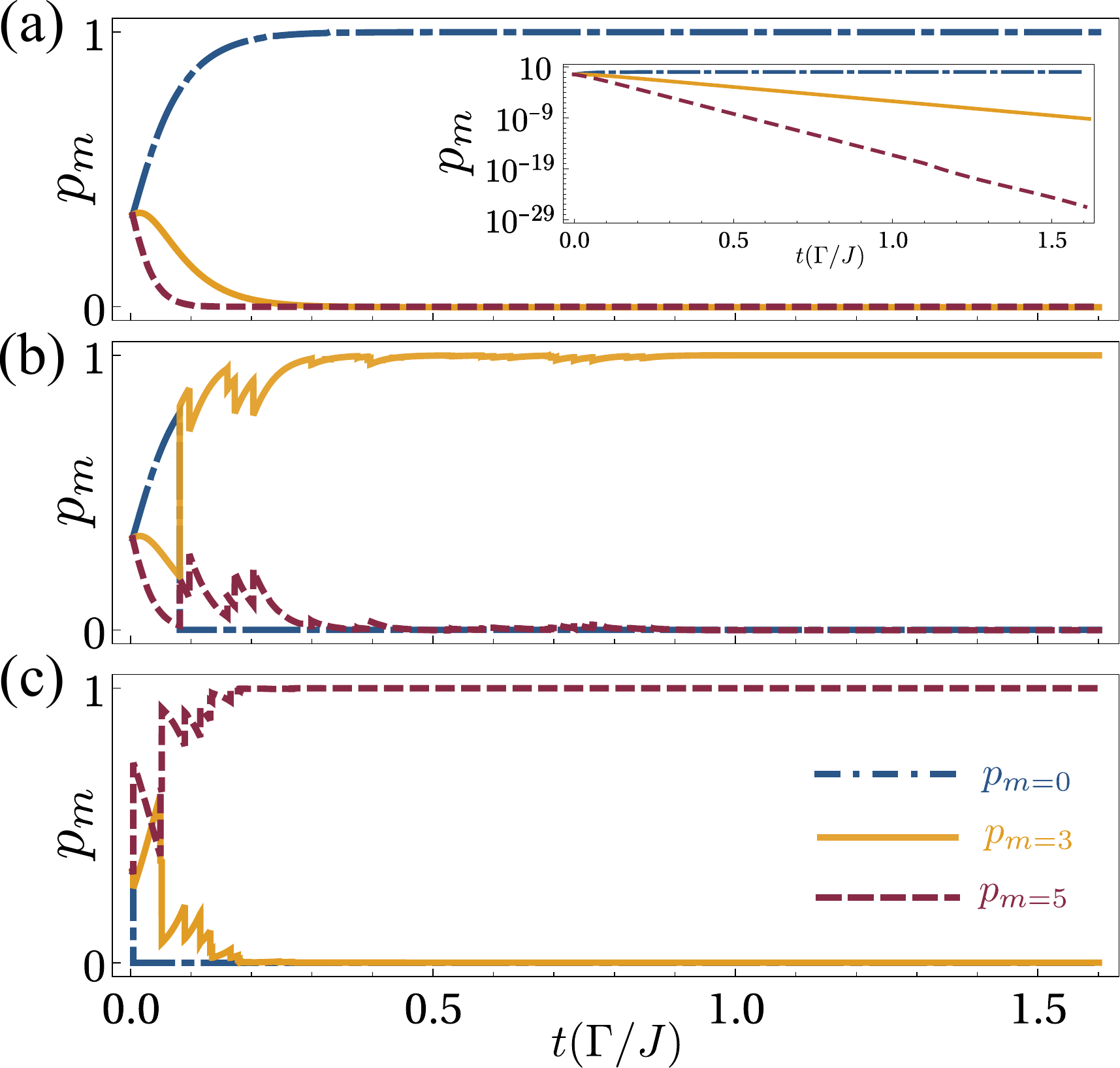}
\end{center}
\caption{Three different quantum trajectories at $\theta=\pi/4$ for the same initial state (a superposition of three eigenstates of $S_x$). Panels (a-c) show the three possible types of trajectories that occur. The inset in (a) shows the exponential decrease of the occupation of non-selected states. Parameters: $J=5$, $\Omega=0.8\Gamma$. }
\label{fig:symmetry}
\end{figure}

The model of squeezed superradiance that we consider here represents, in the particular case $\theta=\pi/4$, one of the simplest implementations of a strong symmetry. The unraveling of the evolution in individual trajectories in this model reveals an effect that we term ``dissipative freezing'' of the dynamics. The phenomenon is depicted in Fig.~\ref{fig:symmetry} (a--c): after initializing the state in a given superposition---in this example, of the $S_x$ eigenstates $|0\rangle$, $|3\rangle$ and $|5\rangle$---the wavefunction eventually evolves into a single one of the eigenstates, with the occupation of any of the other ones decaying exponentially with time. The evolution is thus effectively frozen in one eigenstate for any individual realization of the dynamics, and the conservation law $\dot S_x=0$ is broken.

An eigenstate of a strong symmetry is stationary under this stochastic evolution. To prove this, we consider the general form of any wavefunction undergoing a stochastic, dissipative evolution described by $\tilde H$ and the set of quantum jump operators $\{L_\mu\}$. Starting from an initial state $|\psi(t_0)\rangle$, the wavefunction evolves for a time $t$ experiencing $n$ quantum jumps at times $(t_1,\ldots,t_n)< t$ with jump operators $(L^{(1)},\ldots,L^{(n)})$, where $L^{(i)}\in \{L_\mu \}$. The form of the wavefunction is then given by a nonunitary evolution $|\psi(t)\rangle = \frac{1}{\mathcal N}\tilde U(t,t_n,\ldots,t_0)|\psi(t_0)\rangle$, where $\mathcal N$ is a normalizing constant, and $\tilde U(t,t_n,\ldots,t_0)$ is an  evolution operator given by:
\begin{equation}
\tilde U(t,t_n,\ldots,t_0) = e^{-i \tilde H (t-t_n)}\prod_{m=1}^{n} L^{(m)}e^{-i \tilde H (t_{m}-t_{m-1})},
\end{equation}
with $\prod_{m=0}^n O_m \equiv O_n \cdot O_{n-1}\cdot\ldots \cdot O_0 $. Let us consider a strong symmetry operator $A$, so that $[A,\tilde U]=0$. Therefore, if $|\psi(t_0)\rangle$ is an eigenstate of a strong symmetry $A|\psi(t_0)\rangle=\lambda |\psi(t_0)\rangle$, we obtain
\begin{multline}
A|\psi(t)\rangle = A \tilde U(t,t_n,\ldots,t_0) |\psi(t_0)\rangle \\ =  \tilde U(t,t_n,\ldots,t_0) A|\psi(t_0)\rangle = \lambda |\psi(t)\rangle,
\end{multline} 
i.e. an eigenstate of $A$ remains unchanged at the level of individual trajectories. This proof can be easily extended to the eigenstates of any power $A^n$. This fact may suggest that any quantum trajectory could eventually get ``trapped'' into one of these eigenstates, in a picture somewhat analogue to dark-state cooling~\cite{griessner06a} or population trapping~\cite{aspect88a}.
However, for this to happen, the combination of non-Hermitian Hamiltonian evolution and quantum jumps (which have opposing effects on the occupancy of each eigenstate) should bring the system into one of these eigenstates in the first place. It is a priori not certain that this will occur.

Here, we prove that this is indeed the case when $\dot\rho = -i\Omega[A,\rho] + \Gamma/(2J)\mathcal L_A\{\rho\}$; i.e. dynamics with a single quantum jump $L$ and a general, Hermitian strong-symmetry $A\propto H \propto L$. We set $t_0 = 0 $ and consider an initial state $|\psi(0)\rangle = \sum_m c_m(0)|m\rangle$, expanded in the basis of eigenstates of $A$, $|m\rangle$, with eigenvalue $m$. For \emph{any} general quantum trajectory that evolves for a time $t$ undergoing $n$ quantum jumps, the probability for the final state to be in an eigenstate of $|m\rangle$ takes the form~\cite{arXiv_sanchezmunoz19a}:
\begin{equation}
p(m; t,n) = \frac{1}{\mathcal N} \left(e^{- |m|^2}|m|^{2\alpha}\right)^{t\Gamma/J}|c_m(0)|^2,
\label{eq:S-freezing}
\end{equation}
with ${\alpha = nJ/(t\Gamma)}$ and $\mathcal N$ a normalizing constant. The exponent $t\Gamma/J$ in Eq.~\eqref{eq:S-freezing} tends to enhance the maximum of the function in parenthesis as time increases. Hence, after normalization, $p(m;t,n)$ tends to zero for all $m$ except for the optimum value. Since the function $e^{-x}x^{\alpha}$ has a maximum at $x=\alpha$,  only the eigenstates $|m\rangle$ from the subspace of $A^\dagger A$ yielding the minimum $|\alpha-|m|^2|$ have a non-zero occupancy in the long-time limit  $t\gg J/\Gamma$. Equation~\eqref{eq:S-freezing} thus encapsulates the essence of the dissipative freezing effect and is the main result of this paper: for $t\gg  J/\Gamma$, any general trajectory will be trapped in an eigenspace of $A^\dagger A$, consequently breaking the conservation law $\dot A=0$ if initialized in a superposition of different eigenspaces. In the long-time limit, the total number of jumps recorded in a trajectory allows one to unambiguously determine, from those eigenspaces of $A^\dagger A$ having an overlap with the initial state, which one has the system been trapped into.

\begin{figure}[t!]
\begin{center}
\includegraphics[width=1\columnwidth]{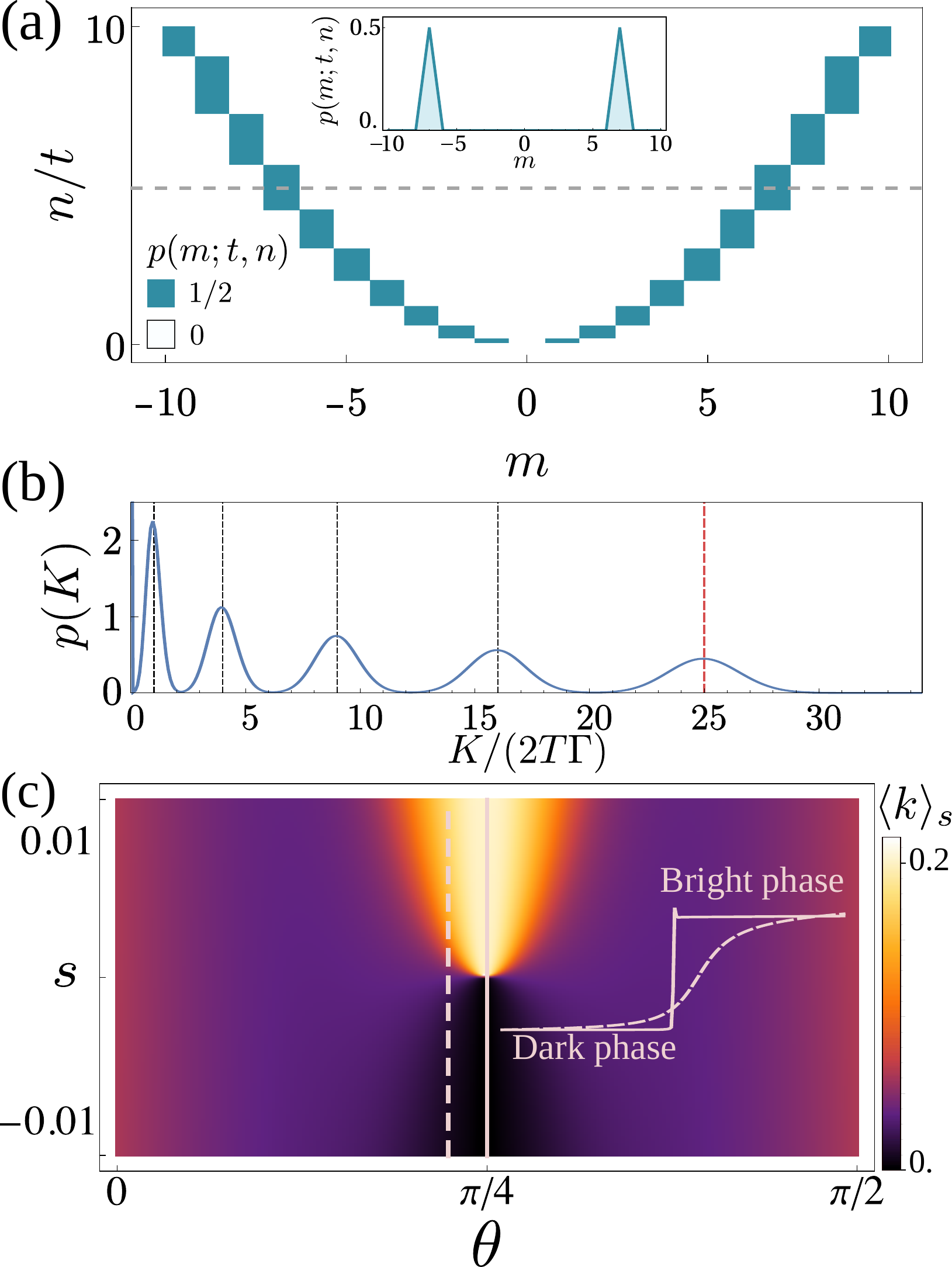}
\end{center}
\caption{(a) Probability distribution for any quantum trajectory at time $t=100 J/\Gamma$, versus the number of jumps $n$, in the model of squeezed superradiance for $\theta=\pi/4$, $J=10$.  The initial state is an equal superposition of all the eigenstates $|m\rangle$ of $S_x$. The resulting wavefunction always freezes into an eigenstate of $S_x^2$. Inset shows the probability distribution for the value $n/t = 5$ indicated by the dashed line. (b) $p_T(K)$ at $T=3\cdot 10^3/\Gamma$ in the case of our model. $N=20$, $\Omega = 0.8\Gamma$, $\theta = \pi/4$ (strong symmetry point). (c) $\langle k\rangle_s$ versus $\theta$, featuring the coexistence between a bright and a dark phase in the vicinity of $\theta=\pi/4$. }
\label{fig:pn_freezing}
\end{figure}

For the particular case that we study in this paper, $H \propto L\propto A=S_x$, $m=-J,\ldots J$. In this case, the eigenstates of $A^\dagger A=S_x^2$ are doubly degenerate. For $t\gg J/\Gamma$, the probability distribution for any quantum trajectory  is $p(m;t,n) \propto\sum_{ m}(\delta_{m,\tilde m}+\delta_{m,-\tilde m})|c_{m}(0)|^2$, with $\tilde m$ the natural number $\leq J$ closest to $\sqrt{nJ/(t\Gamma)}$. The resulting probability distribution versus $n/t$ is plotted in Fig.~\ref{fig:pn_freezing}(a) for an initial state composed of an equal superposition of all the eigenstates.

The phenomenon can be interpreted in terms of a quantum-measurement description of dissipative dynamics~\cite{haroche_book06a,wiseman_book10a,gammelmark14a}. The information provided by the quantum jumps makes the eigenspaces of $A^\dagger A$ with a particular eigenvalue increasingly likely, and continuously updates the state accordingly. We stress that this picture applies to any dissipative dynamics and that this update is different from a projective measurement of $A^\dagger A$ that would collapse the system into one of its eigenstates. In most situations, the update after each jump is not 
able to freeze the state due to the non-Hermitian evolution between jumps, which also changes the occupancy of these eigenstates. 
In our case, this is prevented by the strong symmetry, and the effect of the jumps pile up, giving rise to the phenomenon of dissipative freezing.

\emph{Multimodality}.
Having demonstrated the absence of intermittency in the presence of a strong symmetry, it is important to discuss the implications in other phenomena that are usually also linked to DPTs and dynamical phase coexistence, such as the multimodality of the activity distribution~\cite{ates12a}. 
We analyse the problem from the framework of the thermodynamics of quantum trajectories. The activity is defined as the mean number of quantum jumps undergone by the system per unit time, which can be expressed through the probability distribution $p_T(K)$ of counting $K$ jumps on a time $T$.  In the theory of thermodynamics of quantum trajectories~\cite{garrahan10a,ates12a}, the activity in the long time limit is postulated to follow a large deviation principle $p_T(K) \asymp  e^{-T\varphi(K/T)}$, where the rate function $\varphi(K/T)=-\ln p_T(K)/T$ has the properties of an entropy density~\cite{touchette09a}. Equivalently, the cumulant generating function also has a large deviation form $Z=\langle e^{sK}\rangle \asymp e^{t\lambda(s)}$. Here, $\lambda(s)$ plays the role of a free energy, related to the entropy by a Legendre transformation $\lambda(s) = \max_k[ks-\varphi(k)]$. $\lambda(s)$ encapsulates the statistical properties of the trajectories, and it allows to write the mean activity as $\langle k \rangle \equiv \langle K\rangle/T= \partial \lambda(s)/\partial s|_{s=0}$. $\lambda(s)$ can be obtained as the largest eigenvalue of the tilted Liouvillian $\mathcal W_s(\rho) = \mathcal L(\rho) - (1-e^s)L\rho L^\dagger$, with $L$ the operator inducing the jumps that we are recording. Once $\lambda(s)$ is found this way, the Legendre inverse transformation $\varphi(k)=\max_s[ks-\lambda(s)]$ allows us to obtain $\varphi(k)$. 
This relation, however, requires $\lambda(s)$ to be differentiable for all $s\in \mathbb{R}$ or, equivalently, $\varphi(k)$ to be concave for all $k\in\mathbb{R}$~\cite{touchette09a}. 

In the presence of a strong symmetry, however, $\varphi(k)$ is non-concave and, therefore, $\lambda(s)$ is non-analytic. To show this, we consider a strong symmetry $A$ with eigenstates $|m\rangle$ and only one quantum jump operator, $L=\sqrt{\Gamma/J}A$. For an initial state $\rho(0) = \sum_m c_m(0) |m\rangle\langle m|$, the quantum-jump probability distribution takes the form~\cite{arXiv_sanchezmunoz19a}:
\begin{equation}
p_T(K)= \sum_m \frac{1}{K!}\left(\frac{T\Gamma |m|^2}{J} \right)^K e^{-\Gamma |m|^2 T/J}c_m(0),
\label{eq:pn}
\end{equation}
which is dependent on the initial state via the coefficientes $c_m(0)$. This equation presents the multimodal structure depicted in Fig.~\ref{fig:pn_freezing}(b), plotted for the particular case of our model, $A=S_x$. Each steady state $\rho_0^{(m)}=|m\rangle\langle m|$ with non-zero overlap with $\rho(0)$ manifests as a distinct peak in the counting distribution $p_T(K)$, centered at the value $K_m=T|m|^2\Gamma/J $, which is the emission rate expected for that particular state. The multimodal structure of $p_T(K)$---and consequently of $\varphi(k)$---cannot be obtained from $\lambda(s)$ through an inverse Legendre transformation~\cite{touchette09a}, therefore, it points towards a non-analicity of $\lambda(s)$. As we prove in the Ref~\cite{arXiv_sanchezmunoz19a}, this non-analicity consists of a discontinuity of $\langle k\rangle_s \equiv \partial \lambda(s)/\partial s $ at $s=0$, as it is usually described in the context of dynamical phase transitions~\cite{garrahan07a,garrahan10a,ates12a,flindt13a,hickey14a}. A first-order phase transition in $\langle k\rangle_s$ is therefore linked to the phenomenon of dissipative freezing.

Our model allows us to explore how this discontinuity turns into a continuous crossover as we depart from the strong-symmetry point $\theta=\pi/4$. 
This is shown in Fig.~\ref{fig:pn_freezing}(c), where we plot  $\langle k\rangle_s$ versus $\theta$. For $\theta=\pi/4$, the limit $s\rightarrow 0^+$ features a bright phase characterized by a high activity, whereas for $s\rightarrow 0^-$ we find a dark phase with virtually no quantum jumps (see insets). Away from this point, we obtain a crossover consistent with the first-order phase transition smoothed by finite-size effects  usually observed in finite many-body systems undergoing a DPT~\cite{ates12a}. It implies that $\lambda(s)$ is analytic, and that, in the long-time limit, $p_T(K)$ is unimodal. Unimodality is a consequence of intermittency: the  switching between different dynamical phases destroys the multimodal distribution for times longer than $\tau=-\Re(1/\lambda_2)$~\cite{macieszczak16b}.
Therefore, intermittency is unequivocally connected to a crossover in $\langle k\rangle_s$ at $s=0$, and dissipative freezing, to a discontinuous, first-order transition. Alternatively, dissipative freezing can be described as the survival of multimodality in the long-time limit.

The survival of multimodality is of strong importance in the context of enhanced quantum metrology, where it has been proven that there is a Heisenberg scaling  of the quantum Fisher information  for times shorter than the correlation time $\tau$~\cite{macieszczak16b}. Since systems with a strong symmetry will feature an \emph{asymptotic} quadratic  scaling of the quantum Fisher information for all times, our results may be of  relevance in the design of sensing protocols aimed to exploit this feature on continuous Bayesian parameter estimation from photon counting~\cite{gammelmark14a,kiilerich14a,kiilerich16a}. 
Beyond the model considered here, our results have strong implications for the dynamical characterization of DPTs  in more complex systems where the existence of a strong symmetry can provide a way to tune the Liouvillian gap to zero without  the need of reaching a thermodynamic limit.

\section*{ACKNOWLEDGEMENTS}
C.S.M. kindly acknowledges F. Minganti for fruitful and insightful discussions. B.B and C.S.M are grateful to Juan P. Garrahan for very useful comments and insightful discussions. 
C.S.M. is funded by the Marie Sklodowska-Curie Fellowship QUSON (Project  No. 752180). B.B., J.T. and D.J. acknowledge support from the EPSRC grants No. EP/P009565/1 and EP/K038311/1, and the European Research Council under the European Union’s Seventh Framework Programme (FP7/2007-2013)/ERC Grant Agreement No. 319286 Q-MAC. AGT and DP acknowledge support from CSIC Research Platform on Quantum Technologies PTI-001 and from Spanish project PGC2018-094792-B-100 (MCIU/AEI/FEDER, EU).

\addcontentsline{toc}{chapter}{Bibliography}
\bibliographystyle{mybibstyle}
\bibliography{Sci,arXiv,books}

\end{document}